\documentclass[12pt]{iopart}
\pdfoutput=1
\usepackage{graphicx,iopams}
\begin{document}

\title{Anisotropic thermalization propelled motor}
\author{Hanqing Zhao}
\address{Department of Modern Physics, University of Science and Technology of China, Hefei 230026, China}
\ead{phzhq@mail.ustc.edu.cn}

\begin{abstract}
Brownian motors and Feymann ratchets have been intensively studied in the past decades due to their significance to the foundation of statistical physics. In this work we propose a new type of Brownian motor, i.e., a self-driven motor that only utilizes the temperature difference inside and outside the motor. The motor is a container with asymmetric geometry;  when filling of gas, directional motion occurs if its temperature is different from the environment. The essentially new mechanism is that the asymmetric geometry of the container may induce anisotropic thermalization, which results in a density gradient and propels the motor. The directional motion ceases until the density gradient disappears as the inside temperature approaches the environment temperature.  The same mechanism is also applied to design self-driven  rotators. Possible experimental realizations  are discussed. 
\end{abstract}

\section{\label{sec:1}Introduction}

Since the proposal of Maxwell demon, how to systematically rectify thermal fluctuation has attracted huge attention \cite{maxwell,smo,rec2,feymann,rec1}. It is clear that the Maxwell demon can not exist in an equilibrium system since it is forbidden by the second law of thermodynamics and the detailed balance from statistical mechanics. However, in a non-equilibrium environment, the rectification of thermal fluctuations may take place, e.g., the various Brownian motors \cite{hanggi,hanggi2,reimann,li,li2}, where noisy or periodic signals may result in directional motion of particles in an asymmetric potential \cite{rec2,noisy1,noisy2,noisy3,noisy4}. This phenomenon usually occurs in an environment of uniform temperature. Another class of motors are the Smoluchowski-Feymann ratchets \cite{smo,feymann,exp2,exp1,bang}, where the directional motion or rotation may take place when two connected ratchets with asymmetric geometry are put into higher and lower temperature environments, respectively. Later, Van den Broeck and collaborators simplified the Smoluchowski-Feymann ratchets \cite{broeck1,broeck2}. They designed models with only two asymmetric units [Fig. 1(a)], and found directional motion when putting the two units into environments with different temperatures. By properly arranging them, these models can serve as prototypes of micro-refrigerators and ratchets \cite{broeck3}. Although these works provide solid design principles of a realistic microscopic device that can proceed self-driven directional motion \cite{broeck1,broeck2,broeck3,njp} or rotation \cite{lan,rondoni}, it is unpractical to connect two units rigidly at the microscopic level. Furthermore, to sustain a significant effect, the temperature difference should be big enough, which is unpractical in microscopic level either. To overcome these drawbacks, it has been shown that introducing dissipation into granular ratchets can result in the directional motion of a single-body object \cite{gran1,gran2}, where the violation of the energy conservation is responsible for breaking the time-reversal symmetry \cite{hanggi}. 

In this Letter we propose a self-driven, single-body motor model without dissipation.  The motor is schematically shown in Fig. 1(b).  It is a rigid triangle container with mass $M$. Particles with initial 
temperature $T_1$  and $T_2$  are uniformly distributed inside and outside the container. 
 The key difference from previous motors is that our model involves only a single container filling of particles. We show first by numerical simulation that the directional motion of the container occurs if there is a temperature difference between $T_1$ and $T_2$. This effect exists both for two-dimensional and three-dimensional containers, and either point particles or hard disks give qualitatively same results.  We then illustrate that this effect is caused by a new mechanism what we call anisotropic thermalization. It is different from the mechanism driving the directional motion  of the two-unit Smoluchowski-Feymann ratchets\cite{broeck1}.   Finally, we show that the same mechanism can be applied to design self-driven rotators.

    \begin{figure}
    \centering
    \includegraphics[width=8.9cm]{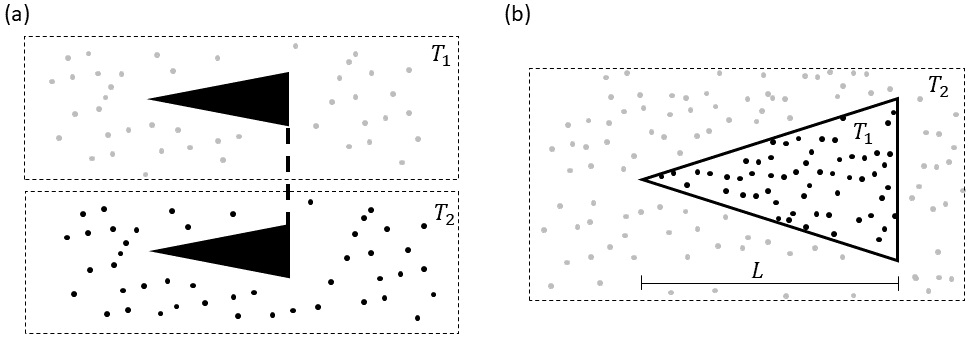}
    \caption{Schematic representation of the motor (a)  proposed by Van den Broeck and collaborators \cite{broeck1} and (b) triangle motor of ours. The former consists of solid bodies putting into environments of different temperatures, while ours are hollow containers filling of gas.  We denote the apex angle in (b) by $\alpha$.}
    \end{figure}

\section{\label{sec:2}Simulations and Results}
\subsection{Self-driven single-body motor} 
We first report the numerical simulation results for the two-dimensional triangular container with point particles inside and outside. The particles interact with the container via elastic collisions with its boundaries. The container can move freely in both horizontal and vertical directions, but is not allowed to rotate. The area of the container is fixed at $S_0=80$ in dimensionless units, and the number of inside particles is $500$ of temperature $T_1$, so that the averaged particle density is unchanged. For a given apex angle $\alpha$ of the triangle, due to the fixed area, its length $L= \sqrt{2S_0/\tan(\alpha/2)}$ [see Fig. 1(b)]. In the following we will use $L$ as the characteristic size of the motor.  The environment is represented by a  rectangle of area $200\times50$ filled with other $500$ particles of temperature $T_2$.  The periodic boundary conditions are adopted. All particles are identical with $m=1$. The mass of container is $M=10$. Initially, the motor is set at rest and the velocities of both the inside and outside particles follow the Maxwell-Boltzmann distribution. The momentum of the system is conserved and set to be zero initially.

 With $\alpha=10^\circ$, Fig. 2(a) shows the probability distribution function (PDF) of the triangle motor center (initially positioned at the origin) at time $t=800$ for $T_{1}=T_{2}=1$ (the Boltzmann constant $k_B$ is set to be unit. With characteristic size $L$ of the motor, the corresponding characteristic time is $\sqrt{m S_0/ k_B T}$. We set this characteristic time as unit of time. We see that the motor center performs a Brownian motion around its initial position. The PDF appears anisotropic along vertical and horizontal axes, indicating that the diffusion constants along the two directions are different. This is a consequence of the fact that the triangle is asymmetric, as in the case of an ellipse particle diffusing in an equilibrium environment \cite{an1,an2}. Therefore, there
is no directional motion on average in an equilibrium situation.

  \begin{figure}
    \centering
    \includegraphics[width=8.9cm]{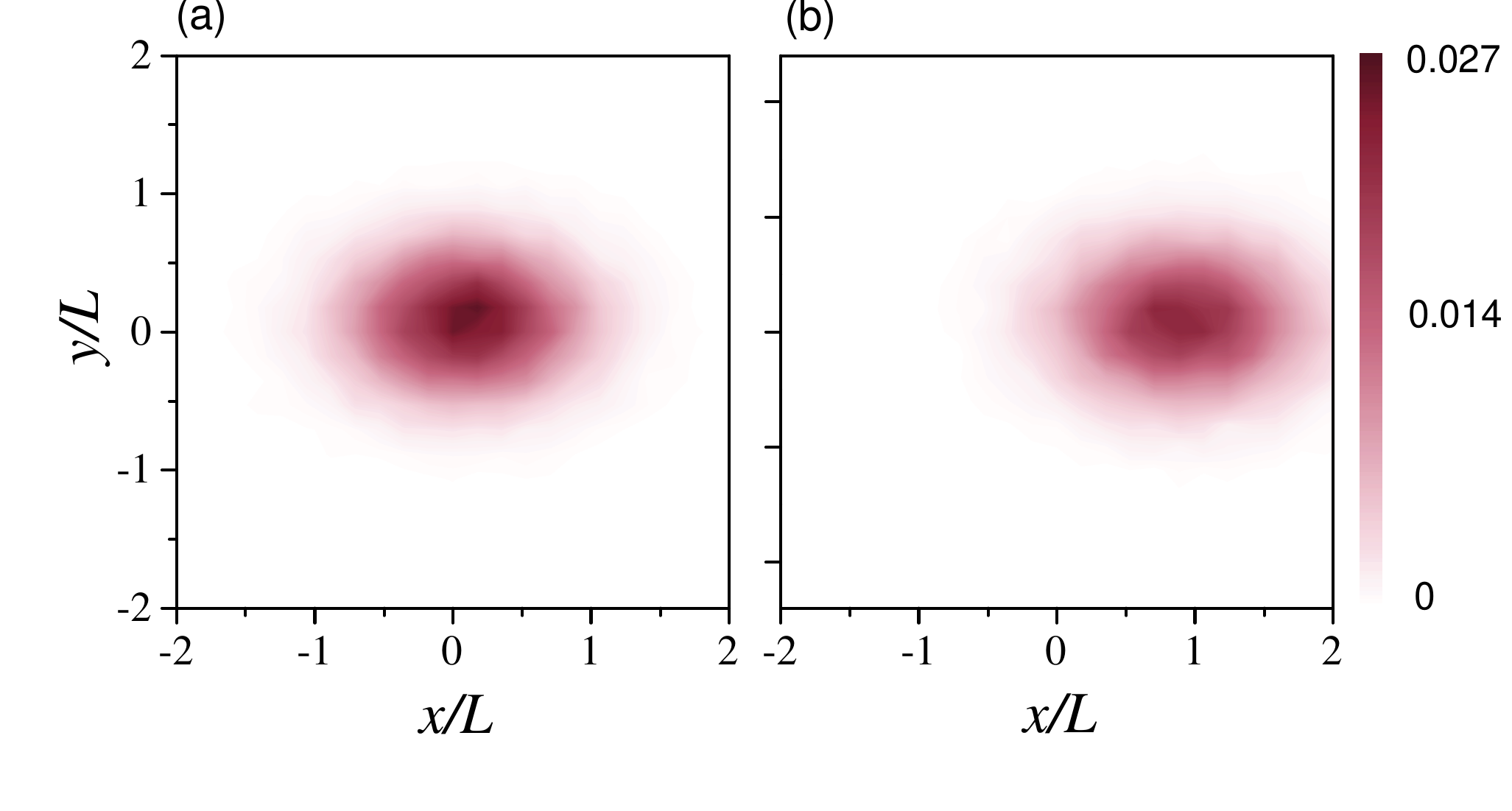}
    \caption{The PDF for the center of mass of the triangle motor in equilibrium initial conditions with $T_1=T_2=1$ (a) and in nonequilibrium initial conditions with $T_1=2$ and $T_2=1$ (b). For both (a) and (b), $\alpha=10^\circ$, $t=800$, and $10^5$ realizations have been taken into account to calculate the PDF. The spatial coordinates are plotted in the unit of motor size $L$.  }
    \end{figure}

Figure 2(b) shows the PDF of the triangle center for $T_{1}=2$ and $T_{2}=1$ at $t=800$. Obviously, the center of the PDF shifts to the right, i.e., a directional motion occurs. Comparing with Fig. 2(a), one can see that it is a combination of the drift motion and the Brownian motion.  There is no directional motion along the vertical direction, suggesting that the directional motion is induced by the asymmetry of the triangle in the horizontal direction.

The directional motion depends on the degree of the asymmetry. Fig. 3(a) shows the average displacement of the triangle center for different $\alpha$ values. We see that $\alpha = 60^\circ$ is a threshold. Below this angle, the motor moves to the right on average, and the maximum displacement is larger when $\alpha$ is smaller. Exceeding this threshold the directional motion  seems to disappear.

    \begin{figure}
    \centering
    \includegraphics[width=15cm]{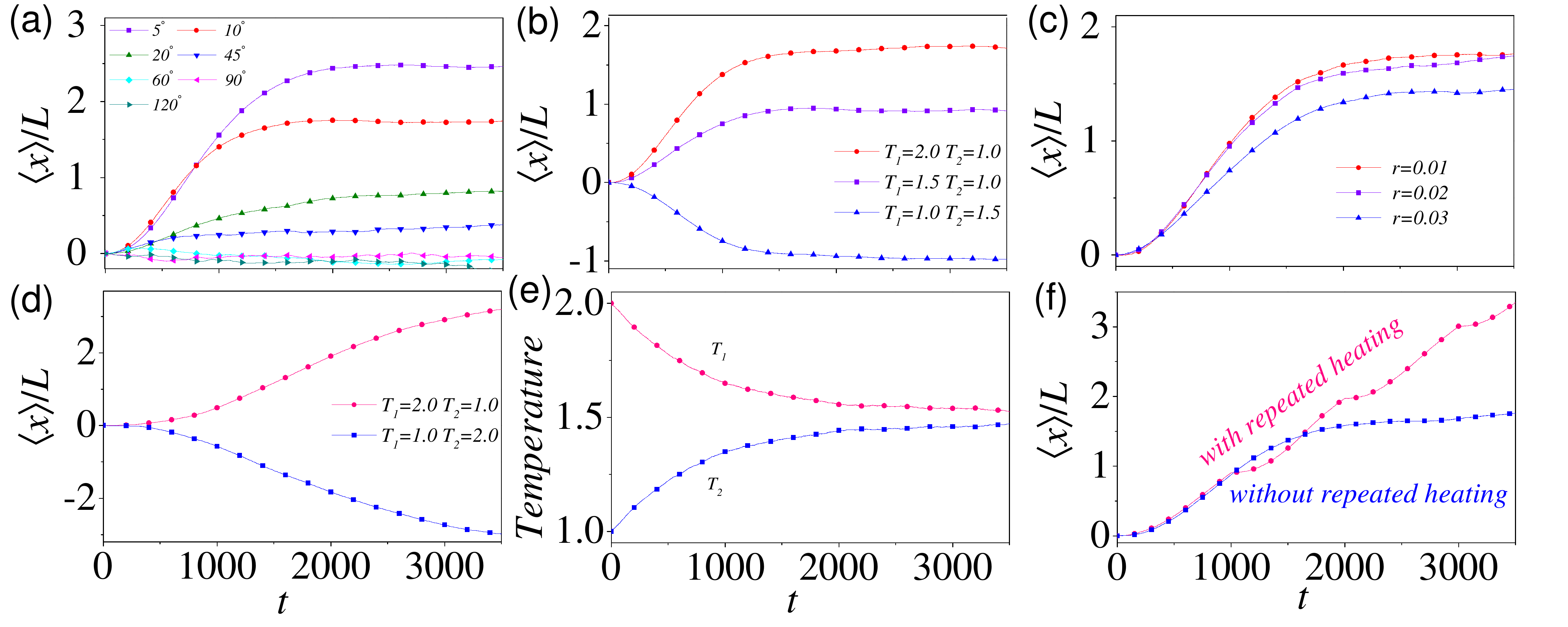}
    \caption{The directional motion of the single-body motor. (a) and (b) show the average displacement of the motor as a function of time for different apex angles and different initial temperatures, respectively. Apex angle and initial temperatures are given in the plots respectively.  (c) shows the average displacement of the motor for different particle radii. (d) shows that the average displacement of the three dimensional motor. (e) shows that the temperature difference decreases with the evolution of time. (f) shows the average displacemnt when considering repeated heating. To perform the repeated heating,  we reset the inside temperature to $T_1 = 2$ after every time interval $\Delta t = 1000$ . In (a) and (c) the initial temperatures are $(T_1, T_2)=(2,1)$, and in (b)、 （c)、 （d） and (f) the apex angles are fixed at $\alpha=10^\circ$. The displacement is represented in unit of motor size $L$. For each data point in all three panels, an ensemble average over $1000$ realizations is taken to evaluate its value.
}
    \end{figure}

The directional motion is also affected by the temperature difference, that is, the higher the temperature difference, the stronger the directional motion will be, as Fig. 3(b) shows. Reversing the temperature difference changes the direction of motion.

The effect remains if we replace point particles with hard disks. The hard disks interact each other by collision. In the study of two-unit Smoluchowski-Feymann motors, it has been shown that the effect shows no qualitative difference when applying the idea gas or the hard disks\cite{broeck1,broeck2}. However, introducing hard disks will increase the computational complexity dramatically. We therefore only use $200$ hard disks with $T_1=2$ to simulate the inside particles, and another $200$ hard disks with $T_2=1$ for outside particles. Fig. 3(c) shows the results for three choices of the hard-disk radii, $0.01$,$0.02$ and $0.03$ respectively. We see that the effect is qualitatively same though it is slightly weakened with the increase of radius.  Based on this fact, we still use the point particles in the following studies.  
   
The effect remains in three-dimensional models.  To show this, we apply a triangular box with a thickness of $2$ be the container, and the environment is represented by a cuboid of $200\times50\times4$. Fig. 3(d) shows that with the same number of particles and the temperature difference, the effect still exist. This result is similar to the case of two-unit Smoluchowski-Feymann motors, where it has been shown that the effect remains for 3D motors. 

We see from these figures that the shift from the initial position saturates eventually. Fig. 3(e) shows the inside and outside temperature versus the evolution time. The two temperatures approach to an equilibrium value which correspond to the saturation. The environment temperature rises since only a finite number of particles are used to represent the environment. Therefore, to maintain a continuous directional motion, we have to repeatedly heat the motor, as Fig. 3(f) shows. 

\subsection{Anisotropic thermalization}

Next, we discuss the propelling mechanism of the motor.  We first show that the mechanism driving the two-unit Smoluchowski-Feymann ratchets should not be the dominant mechanism for our motor. We note that our motor is mathematically equivalent to one of the two-unit Smoluchowski-Feymann ratchet proposed by Van den Broeck et. al. \cite{broeck1} if assuming that the inside and outside particles keep uniform density and ignore the correlated collisions. The analytical approach developed in Refs. \cite{broeck1,broeck2} can be straightforwardly applied to our model with these assumptions.  When restricting the motion on horizontal axis, we obtain the stationary speed (see supplementary material section 1):
\begin{equation}
\langle V\rangle=\rho_{1}\rho_{2}(1-\sin\frac{\alpha}{2})\sqrt{\frac{m\pi k_B}{8M^2}} \frac{(T_{1}-T_{2})(\sqrt{T_{1}}+\sqrt{T_{2}})} {(\rho_{1}\sqrt{T_{1}}+\rho_{2}\sqrt{T_{2}})^2}.
\end{equation}
This formula is identical to the two inverted triangles model \cite{broeck2}. This is not surprising since the sum of the integral with respect to the two triangles in that model is equivalent to the integral with respect to inside and outside regions of the single triangle in our model. 

However, numerical investigations suggest that Eq. (1) fails in predicting the effect we observed.  As an example, for the case of $\alpha=20^\circ$, we have $\langle V\rangle \sim 5\times 10^{-4}$ by Eq. (1), but our simulation result [see Fig. 3(a)] suggests that ${\rm d}\langle X\rangle/{\rm d}t\sim 8.1\times 10^{-3}$ before saturation, which is one order larger than the former. Considering that Fig. 3(a) is for the 2D motion while Eq. (1) is applicable for 1D, we re-perform the simulation by restricting our motor to move along the horizontal direction. It gives  ${\rm d}\langle X\rangle/{\rm d}t\sim5.4\times 10^{-3}$, which is of the same order of the 2D simulation. 

   \begin{figure}
    \centering
    \includegraphics[width=7.9cm]{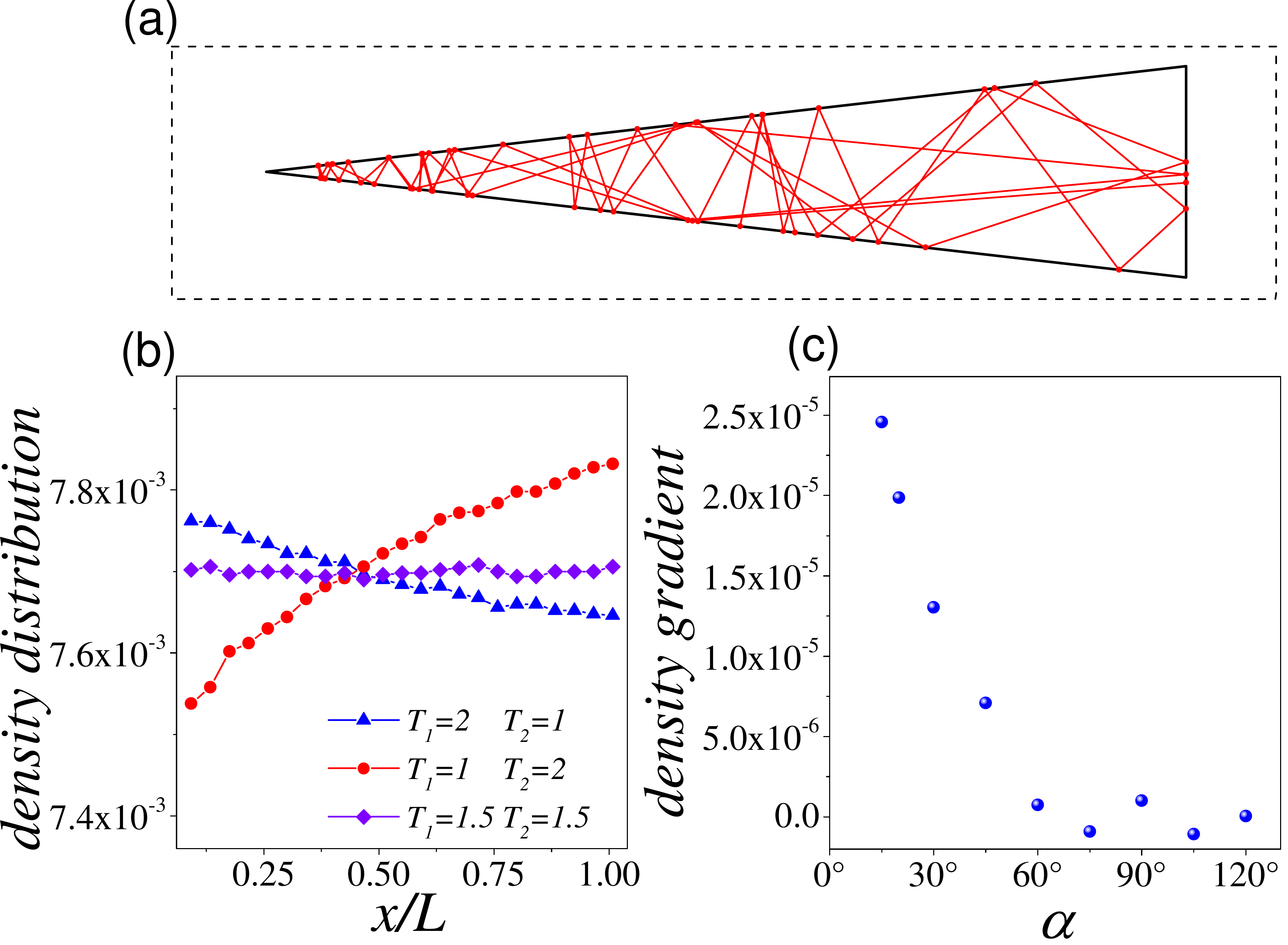}
    \caption{Illustration of anisotropic thermalization. (a) A segmental trajectory of a particle in a triangle. (b) The density distribution along the symmetry axis of the triangle with $\alpha=10^\circ$. The initial temperatures are  given in the plot. (c) The density gradient versus $\alpha$. The density distribution is measured at $t=150$ and an ensemble average over $10^5$ realizations has been used for obtaining the results.
}
    \end{figure}

We then verify that the observed directional motion is resulted by the anisotropic thermalization. It has been shown that with elastic collisions directional motion can not take place for a single solid body even if it has an asymmetric shape \cite{gran1,gran2}. But for containers filled with gas, the situation changes. The significant difference is that the inside particles may change the local density due to anisotropic thermalization.  Since collisions with the container frame is elastic, the particles only change their velocity along the normal direction at the collision point. For a triangle with a sufficient small apex angle, particles in the apex-angle region will collide with the legs more frequently, see Fig. 4(a).  As a result, though a collision with the bottom may exchange a bigger amount of horizontal momentum than that with a leg, the momentum exchange with the environment can be faster in the apex-angle region than in the bottom region on average. This effect induces,  in the case of  $T_1>T_2$, that particles accumulate in the  apex-angle region, which results in a density gradient along the horizontal direction as Fig. 4(b) shows. The density gradient establishes very fast,  corresponding to the period in Fig. 3 that is close to the origin in which  $\langle X \rangle$ increases slowly (see supplementary material section 2). Once the density gradient is established, the anisotropic thermalization process finishes. 

The master equation approach fails in our model since the inside particle density gradient invalidates the asumpation of uniform density.  On the other hand, the density gradient provides a new mechanism of directional motion. As the temperature difference decreases (due to the exchange of the energy and momentum between the inside and outside particles),  the particles at the high-desity side will relax to the low-density side. When these particles hit the boundary of the container at the low density end, a net momentum is transferred to the container, and results in the directional motion. The directional motion ceases untill the gradient vanishes
\begin{figure*}
\centering
\centerline{\includegraphics[width=17cm]{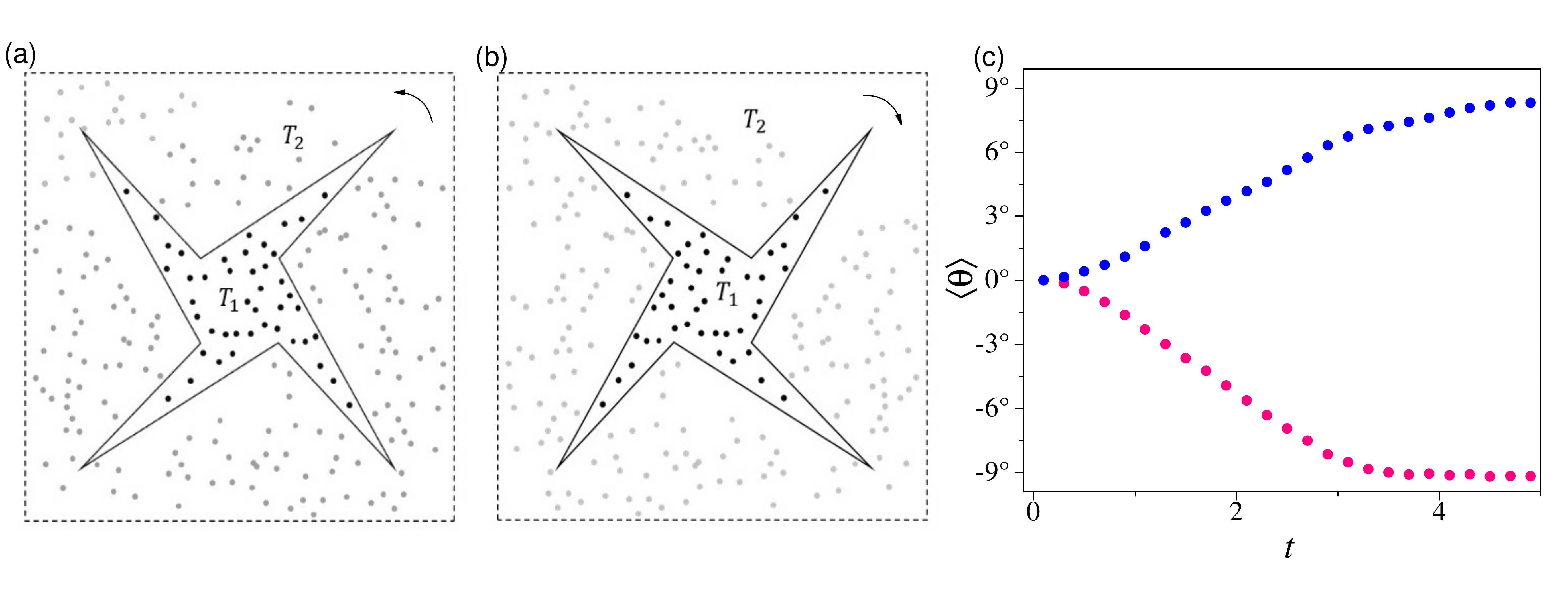}}
    \caption{The schematic illustration of a rotating motor (a) and its mirror (b).  In (c), we plot the simulated average rotation
angle as a function of time for the motor (up circles) and its mirror (bottom circles), respectively. The initial temperatures are $(T_1,T_2 )=(2,1)$. The numbers of particles, inside and outside the motor, are both $500$.
}

\end{figure*}
 
In Fig. 4(c), we plot the gradient at $t=150$ as a function of apex angle. We find that it has a strong correlation with Fig 3(a) : For $\alpha\geq 60^\circ$ the gradient disappears. It can be understood that at $\alpha=60^\circ$ the directional motion must disappear.  For an equilateral triangle, since the system has a rotation symmetry of $2\pi/3$, if it moves right, then it will move equally to the $2\pi/3$ and $4\pi/3$ direction, this cancels the movement in any direction, thus directional motion can not occur.  However, though simulation results indicate so, we have no analytical justification whether the density gradient exactly vanishes with $\alpha>60^\circ$.

\subsection{Self-driven single-body rotator}

Finally, the anisotropic thermalization mechanism can be exploited to design rotating motors as well. Figure 5(a) shows
such a motor, where the motor is not allowed to move translationally but rotate around its center freely.   We perform simulations with $500$ particles of $T_1=2$ inside the motor, and other $500$ particles of $T_2=1$ outside the motor. The moment of the inertia of the container is set to be $10$.  Fig. 5(c) shows the directional rotation occurs counterclockwisely.  By a mirror transformation with respect to the vertical axis we obtain a mirror motor [Fig. 5(b)]. Our simulation shows [Fig. 5(c)] that it rotates clockwisely.  It is therefore the symmetry that induces the directional rotation. 
It should be noted that, there are other works of rotating ratchets \cite{pna1,pna2}, but they are different from our design. In particular, those motors are not heat engines; they exploit the asymmetrical design of the ratchet and are driven by bacteria without temperature difference.

\section{\label{sec:3} Conclusion and discussion}

In summary, filling gas to a proper asymmetric container can induce directional motion if the gas temperature is different from the environment. We show that the directional motion of
the motor resulted from anisotropic thermalization. Though the mechanism propelling the two-unit Smoluchowski-Feymann ratchets may also approximately exist in our model, it is negligible compared to the
effect of the anisotropic thermalization.  Indeed, we have shown that with the same temperature difference the directional speed of our single-body motor is one order higher than the corresponding two-unit motor.  We therefore reveal a new mechanism for directional motion of Brownian motors.

In our study the rotational freedom is excluded in the triangle motor. Without this restriction the numerical simulations may become more complicated. Taking into account of rotation deserves further studies. Nevertheless, we can estimate the possible consequences. In general, with the rotational
freedom the directional motion may be destroyed in a long time. In this case, the directional motion may appear as an enhanced effect to the diffusion process. In a short time, the directional motion may remain for the triangle motor with small enough apex angle and has a long extension along the horizontal direction, where the rotation can be suppressed due to the big moment of inertia.

In principle, our model may be more accessible to experimental implementations. 
For the sake of principle illustrating, designing a device (a shift container or a rotator container) at the macroscopic level, employing hard spheres or magnetic colloidal particles to mimic the gas \cite{mag,gra}, may be favorable. In such a device, the simulation condition may be applied directly to the real experiment. 

A microscopic realization is also possible, we do not need an extreme temperature difference, dilute gas medium, or a light enough container.  We have checked that the maximum displacement of directional motion behaves as $\sim \Delta T/T$.  We have also checked that simultaneously increasing inside and outside particles as well as the mass of the container to the fixed ratios among their total masses, the maximum shift keeps almost unchanged. (For detailed illustrations please see the supplementary materials, section 2) The extension to high density gas thus will not influence the effect.  With these facts, one can estimate that in normal temperature environment ($T=300K$), a pollen-size motor ($\sim10\mu m$) can move a net distance of $1/10$ motor size if there is a temperature difference of $\Delta T=10K$. It seems relatively weak. However, any  motor needs continuous supply of energy.  As we have shown that repeated heating the motor can sustain a continuous directional motion.  
Recently, photosensitizers of nano particles which can be excited sensitively by infrared light have been produced. They have been applied to raise the temperature of living tumors by distributing them around the tumors \cite{yang}. This technique can be applied to raise the temperature of a microscopic container filled with photosensitizers, even when the container has micrometer size.  By continuous heating the container one can sustain the temperature difference, so that the motor can be propelled to move persistently and therefore its directional motion can be detected.

\section*{Acknowledgment}
We are grateful to W. Wang and Y. Lan for useful discussions. We acknowledge support by NSFC (Grant No. 11535011 and 11775210).

\section*{References}

\bibliography{Reference}

\end{document}